\documentclass[twocolumn]{aastex62}
\usepackage{threeparttable}

\received{2018 September 24}
\revised{2018 November 30}
\accepted{2018 November 30}

\shorttitle{Transmission Spectroscopy of Warm Saturns}
\shortauthors{Deibert et al.}

\begin{document}

\title{HIGH-RESOLUTION TRANSIT SPECTROSCOPY OF WARM SATURNS}

\correspondingauthor{Emily K. Deibert}
\email{deibert@astro.utoronto.ca}

\author{Emily K. Deibert}
\affil{Department of Astronomy \& Astrophysics, University of Toronto, 50 St. George St., Toronto, ON M5S 3H4, Canada}

\author{Ernst J. W. de Mooij}
\affiliation{School of Physical Sciences and Centre for Astrophysics \& Relativity, Dublin City University, Glasnevin, Dublin, Ireland}

\author{Ray Jayawardhana}
\affiliation{Department of Astronomy, Cornell University, Ithaca, New York 14853, USA}
\author{Jonathan J. Fortney}
\affiliation{Department of Astronomy \& Astrophysics, University of California Santa Cruz, Santa Cruz, CA 95064, USA}
\author{Matteo Brogi}
\affiliation{Department of Physics, University of Warwick, Coventry CV4 7AL, UK}
\author{Zafar Rustamkulov}
\affiliation{Department of Astronomy \& Astrophysics, University of California Santa Cruz, Santa Cruz, CA 95064, USA}
\author{Motohide Tamura}
\affiliation{Department of Astronomy, The University of Tokyo, 7-3-1 Hongo, Bunkyo-ku, Tokyo 113-0033, Japan}
\affiliation{National Astronomical Observatory of Japan, NINS, 2-21-1 Osawa, Mitaka, Tokyo 181-8588, Japan}
\affiliation{Astrobiology Center, NINS, 2-21-1 Osawa, Mitaka, Tokyo 181-8588, Japan}

\begin{abstract}
We present high-resolution optical transmission spectroscopy of two sub-Saturn mass transiting exoplanets, HAT-P-12b and WASP-69b. With relatively low densities and high atmospheric scale heights, these planets are particularly well-suited to characterization through transit spectroscopy, and serve as ideal candidates for extending previously-tested methods to lower planetary masses. Using a single transit for each planet, we take advantage of the Doppler cross-correlation technique to search for sodium, potassium, and water absorption features. Our analysis reveals a likely ($3.2\sigma$) detection of sodium absorption features in the atmosphere of HAT-P-12b, and enables us to place constraints on the presence of alkaline and molecular species in the atmospheres of both planets. With our results, we highlight the efficacy of ground-based campaigns for characterizing exoplanetary atmospheres, and pave the way for future analyses of low-mass planets.
\end{abstract}


\section{Introduction} \label{sec:intro}
While thousands of transiting exoplanets have been discovered, relatively little is known about their atmospheric properties and compositions. The main limitation in characterizing a planet's atmosphere lies in the extreme difference in brightness between the planet and its host star, typically preventing direct atmospheric emission from being detected. However, recent progress in the field has made use of transmission spectroscopy during transits, when the light from the host star passes through the exoplanet's atmosphere and allows for the detection of atomic or molecular species.

\cite{Charbonneau02} used transmission spectroscopy to find evidence of sodium in the atmosphere of the hot Jupiter HD 209458b, marking the first detection of an atmosphere around a planet outside our solar system; this observation was later confirmed by \cite{Snellen08}. Other notable contributions include detections of hydrogen, oxygen, and carbon in an evaporating planetary exosphere (\citealt{Vidal-Madjar03, Vidal-Madjar04}), as well as detections of carbon monoxide (\citealt{Snellen10}) and potassium (\citealt{Sing11}). Analyses of transmission spectra have also led to a number of discoveries of molecular compounds in exoplanetary atmospheres, including in particular various water vapour detections (e.g. \citealt{Deming13} and \citealt{Mandell13}, among others). More recent work has focused on the detection of atmospheric clouds and hazes (e.g. \citealt{Pont13}, \citealt{Nikolov15}, and \citealt{Sing16}, among others). Despite a great deal of progress in the study of exoplanetary atmospheres, however, transmission spectroscopy has to date almost exclusively targeted high-mass hot Jupiters at relatively low spectral resolutions, with a few exceptions such as the warm Neptune GJ 436b (e.g., \citealt{Lothringer18}), and the super-Earths GJ 1214b and 55Cnc e
(\citealt{deMooij13}, \citealt{deMooij14}, and \citealt{Kreidberg14}, among others). In the case of 55Cnc e,  \citealt{Esteves17} also conducted a high-spectral-resolution search for water vapor.

The present paper focuses on characterizing the atmospheres of HAT-P-12b (\citealt{Hartman09}) and WASP-69b (\citealt{Anderson14}), two sub-Saturn mass transiting exoplanets with equilibrium temperatures of $\sim$ 1000 K and strongly inflated radii of $\sim 0.96 R_J$ and $\sim 1.06 R_J$ respectively, resulting in large atmospheric scale heights corresponding to $\sim$ 1 \% of their radii. The atmospheres of both exoplanets have previously been observed, resulting in findings that will guide our analysis. In the case of HAT-P-12b, \citealt{Line13} made use of transmission spectra obtained using the Wide Field Camera 3 (WFC3) aboard the Hubble Space Telescope (HST) to show a lack of expected water absorption features in its atmosphere, suggesting the presence of high-altitude clouds.  \cite{Sing16} followed up with the Space Telescope Imaging Spectrograph (STIS) to observe HAT-P-12b in the full optical range, and found evidence of cloud or haze aerosols as well as potassium. In the case of WASP-69b, \cite{Casasayas17} analyzed two transits observed with the High Accuracy Radial velocity Planet Searcher (HARPS-North) spectrograph and reported a $5\sigma$ detection of atmospheric sodium absorption in the D${}_2$ line of the sodium doublet, but not in the D${}_1$ line.

\begin{table*}[htbp!]
\centering
\caption{Summary of Observations}
\begin{tabular}{c  c  c  c  c  c  c}
\footnotesize Planet & \footnotesize Date (UT) & \footnotesize Instrument/Telescope & \footnotesize Duration (hrs) & \footnotesize No. of Frames & \footnotesize Exp. Time (s) & \footnotesize Avg. S/N per Spectral Pixel \\
\toprule
\footnotesize HAT-P-12b & \footnotesize May 5, 2017 & \footnotesize HDS/Subaru & 8.4 & 84 & 300 & 0.0043  \\
\footnotesize WASP-69b & \footnotesize June 10, 2017 & \footnotesize GRACES/Gemini & 5.2 & 107 & 140 & 0.0014  \\
\hline
\label{tab:obs}
\end{tabular}
\end{table*}

\subsection*{}
Here we present high-resolution spectroscopy of HAT-P-12b and WASP-69b, focusing on sodium and potassium lines as well as water absorption features. Our paper is structured as follows. In Section \ref{sec:obs} we describe the observations obtained for each planet, and in Section \ref{sec:reduc} we describe data reduction techniques used to correct for various systematic effects. We present our analysis in Section \ref{sec:analysis}, and comment on our results in Section \ref{sec:discussion}. Our concluding remarks are provided in Section \ref{sec:conclusion}, and an Appendix containing further details on our techniques is included.

\section{Observations} \label{sec:obs}
We observed HAT-P-12b using the Echelle High-Dispersion Spectrograph (HDS; \citealt{Noguchi02}) on the Subaru telescope. The observations were taken during a period of 8.4 hours on May 5th, 2017 UT, and had a typical spectral resolution of 80,000. The wavelength coverage of the observations was 538--799 nm, with a gap from 658--680 nm between the two CCDs of the spectrograph. The observations were taken with the StdRb observing mode\footnote{\url{https://www.naoj.org/cgi-bin/hds_efs.cgi}}
and the \# 2 image slicer, which allowed for an extremely high resolution but prevented us from removing sky emission lines directly from the spectra (for further details on our sky emission reduction, see Section \ref{sec:reduc}). 

Each observation was made with an exposure time of 300 s. The total 8.4-hour observation period represents one transit of HAT-P-12b, with 24 frames during the exoplanet's ingress, transit, and egress. 

WASP-69b was observed using the Gemini Remote Access to the Canada France Hawaii Telescope (CFHT) ESPaDOnS Spectrograph (GRACES; \citealt{Chene14}), which uses the Gemini North telescope in tandem with the Echelle SpectroPolarimetric Device for the Observation of Stars (ESPaDOns; \citealt{Donati03}) at the CFHT. The observations were taken during a period of approximately 5.2 hours on June 10th, 2017 UT. The wavelength coverage of the data was 395--1044 nm, but due to low fiber throughput the data were only useful at wavelengths of 420--1010 nm, with maximum sensitivy between 490--950 nm. We therefore focused on these wavelengths, and achieved a typical spectral resolution of $\sim 60,000$. Our observations made use of the Star-Only (4 slice) mode of the spectrograph, which uses a single fiber to yield extremely high spectral resolution but, again, prevented us from removing sky background emission. The reduction process is detailed in Section \ref{sec:reduc}.

Each observation was made with an exposure time of 140 s. The full observation represents one transit of the exoplanet, with 58 out of 107 frames during the ingress, transit, and egress.

Our observations are summarized in Table \ref{tab:obs}.

\section{Data Reduction}
\label{sec:reduc}
The data were initially reduced using the HDSQL\footnote{https://www.naoj.org/Observing/Instruments/HDS/hdsql-e.html} pipeline available at the Subaru observatory (\citealt{Noguchi02}) for HAT-P-12b, and the Open source Pipeline for ESPaDOnS Reduction and Analysis (OPERA; \citealt{Martioli12}) for WASP-69b. Following this step, we interpolated all spectra from each set of observations to a common wavelength grid in the telluric rest frame using a cubic spline in order to facilitate our analysis.

Contaminating cosmic rays were removed through median filtering. The data were binned and a threshold of 5 median absolute deviations was applied, with points falling above this threshold flagged as cosmic rays and discarded from further analysis.

We then applied a correction to adjust for the blaze function, which is a grating-dependent variation in brightness at different orders in the data introduced by the use of an Echelle spectrograph. This effect was corrected for on a per-order basis for both sets of observations. We accomplished this by first dividing out a low-order polynomial fit to the ratio of the science spectrum with respect to a reference spectrum---in our case, the first spectrum of the dataset---from each individual frame. The divided spectra were then binned and fit with a polynomial, which was evaluated at the wavelength range of the original, unbinned spectra, and divided out of the original spectra in order to remove the effects of the blaze function. The results can be seen in Figs. \ref{fig:datareduc_12} and \ref{fig:datareduc} in Appendix \ref{app:datareduc}, where the wavelength-dependent effects of the blaze response in the first panels have been corrected for in the subsequent panels.

\subsection{Removal of Telluric Emission Lines}
For both planets, strong telluric sodium emission features were visible throughout the observations. As mentioned in Section \ref{sec:obs}, the image slicers used in our observations prevented us from obtaining separate sky observations, and we thus had to manually remove sky emission lines from our data.

We note that the strength of the sodium emission features in our data seemed to vary in accordance with the change in airmass during the observations. Furthermore, contaminating emission features were \textit{not} present in the spectral absorption lines of potassium, which is the other element targeted in this analysis.

\begin{figure}
\centering
\includegraphics[width=\linewidth]{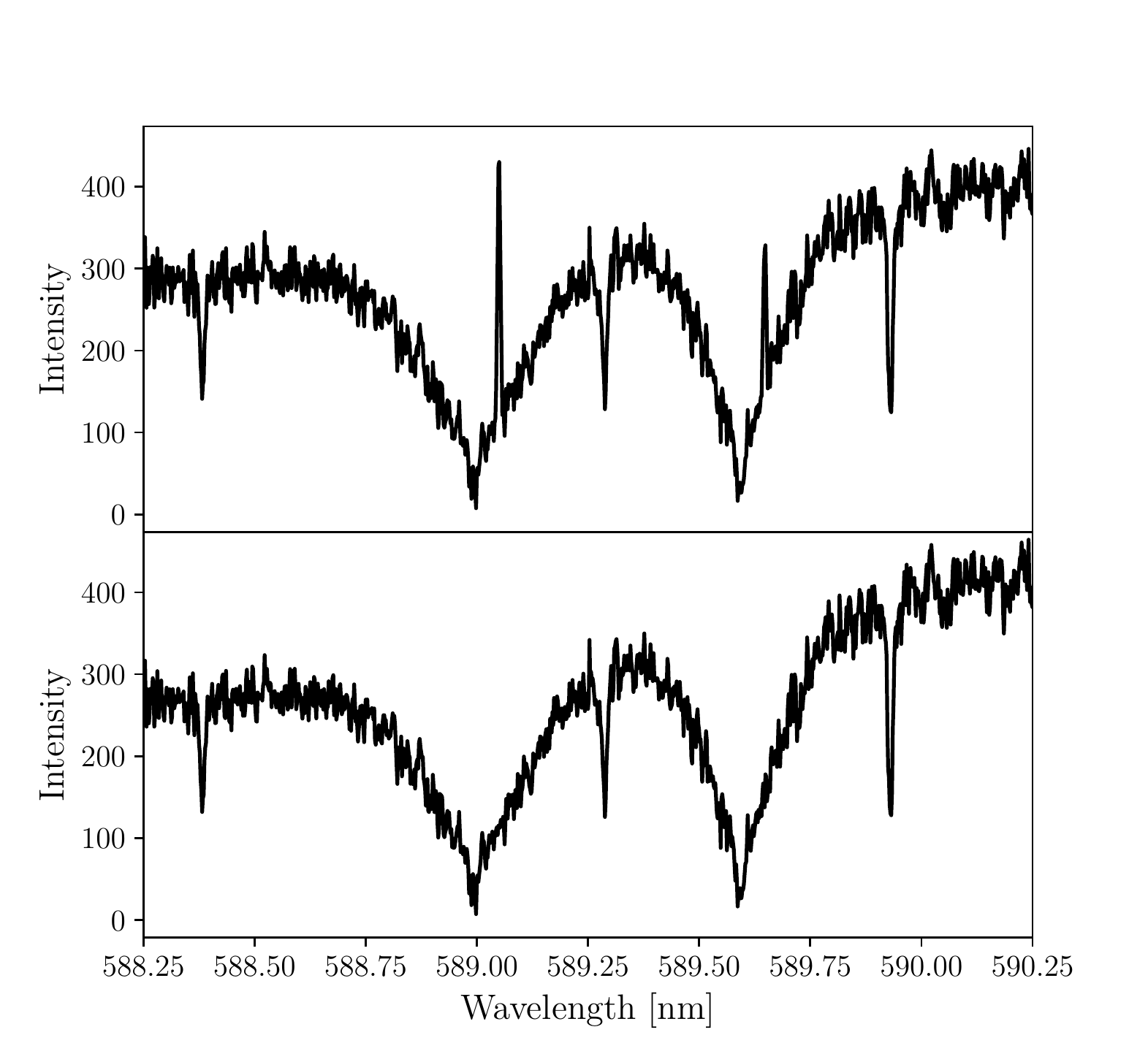}
\caption{The correction for telluric sodium emission as applied to one HAT-P-12b observation. The top panel shows the data before the correction, with emission features present in both lines of the sodium doublet, while the bottom panel shows the same observation after the features have been removed.}
\label{fig:tellna_12}
\end{figure}

To correct for these emission features, we fit them with the sum of a Gaussian and polynomial function, and subtract off the Gaussian in order to preserve the noise in the data. We found that a first-order polynomial was sufficient in characterizing the shape of the emission features, as they were present only in a very small region of the sodium doublet that could be approximated as linear (see Fig. \ref{fig:tellna_12}). The emission features are removed in this way from both lines of the sodium doublet for every spectrum of both the HAT-P-12b and WASP-69b observations. 

Fig. \ref{fig:tellna_12} shows a comparison of one frame of our HAT-P-12b observations before and after the telluric sodium emission has been corrected for. Similar features, though with varying strengths, were observed throughout both sets of observations.

\subsection{Correction of Systematic Effects}
\label{subsec:sysrem}
After initial corrections, the spectra are largely dominated by contaminating telluric and stellar absorption lines (see Fig. \ref{fig:datareduc_wasp69}, and Figs. \ref{fig:datareduc_12} and \ref{fig:datareduc} in Appendix \ref{app:datareduc}).  To correct for these, we make use of the {\sc Sysrem} algorithm (\citealt{Tamuz05}), which was originally designed to correct systematic effects in a large set of photometric light curves. Following the example of previous atmospheric characterization work, however (e.g. \citealt{Birkby13, Esteves17}), we take advantage of the fact that {\sc Sysrem} is well-suited to removing systematic effects that appear in many datasets (in our case, spectra) of the same sample. Due to the large, rapidly-changing radial velocities of the exoplanets during their transits, {\sc Sysrem} can be used to remove the telluric and stellar absorption features, which are stable in time, without affecting the signal from the exoplanets' atmospheres. We invite the reader to consult \cite{Tamuz05} for full details of the {\sc Sysrem} algorithm.

\begin{figure}
\centering
\includegraphics[width=\linewidth]{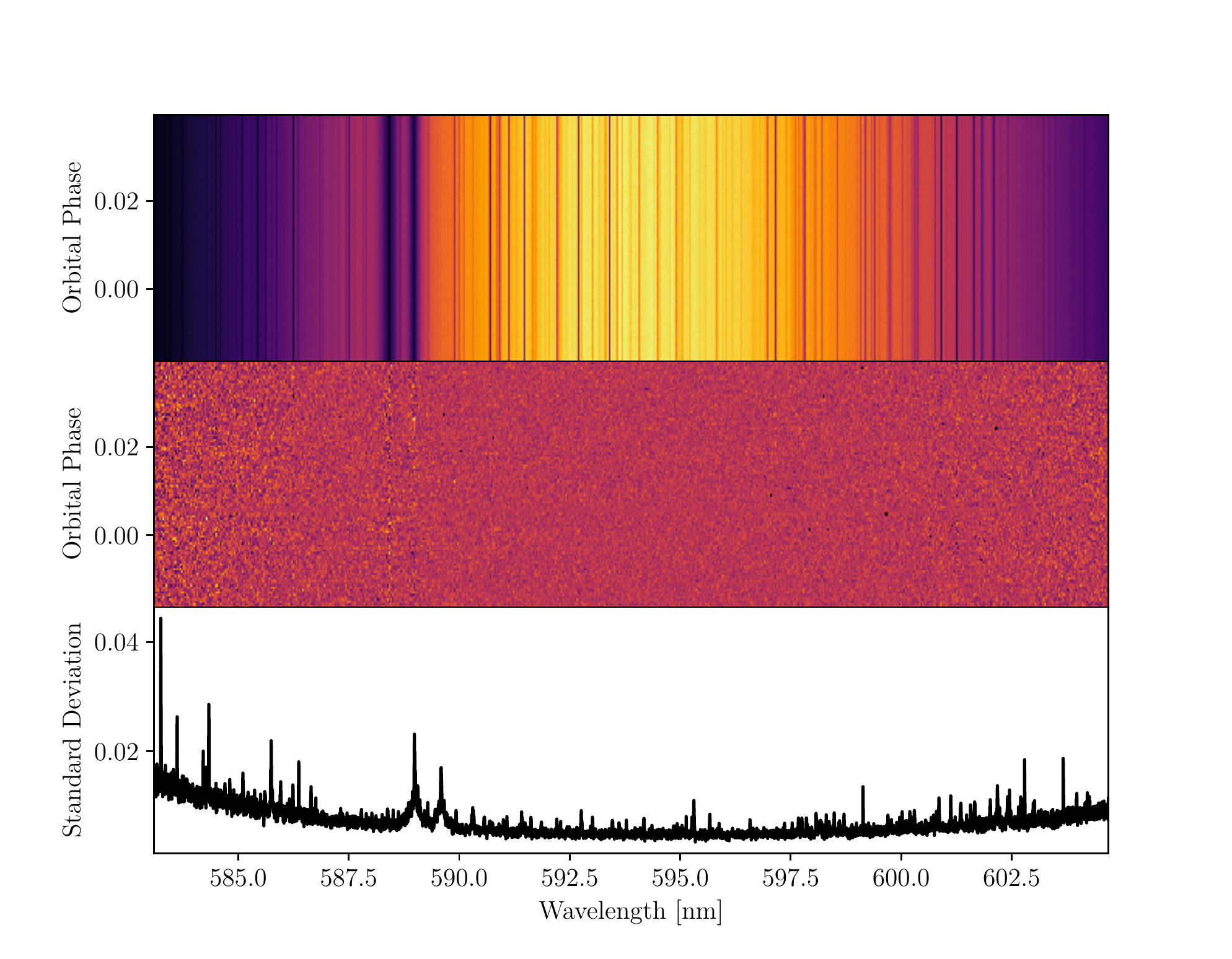}
\caption{The typical data reduction process used in our analysis, for a WASP-69b spectrum. The top panel shows the data in one order after the blaze correction is applied and cosmic rays are flagged. The middle panel shows the results of the {\sc Sysrem} algorithm, and the bottom panel shows the standard deviation of each wavelength band after {\sc Sysrem}.}
\label{fig:datareduc_wasp69}
\end{figure} 

We took the average airmass measured during each exposure as a first approximation to the systematic trends to be removed by the {\sc Sysrem} algorithm. To account for possible variations in the data between separate orders, {\sc Sysrem} was applied to each order individually. We tested various numbers of iterations of the algorithm and eventually chose to apply 6 iterations to each order, after finding that this was sufficient for removing the majority of systematic effects present in the data, minimizing the RMS per order, and preventing overfitting. Figs. \ref{fig:datareduc_12} and \ref{fig:datareduc} in Appendix \ref{app:datareduc} show the results of this process as applied to all orders, while Fig. \ref{fig:datareduc_wasp69} shows a specific example of the reduction process as applied to the order containing the sodium doublet in our WASP-69b observations.

We note that the algorithm performs poorly around strong lines. To account for this, we followed \cite{Snellen10} and \cite{Esteves17} and weighted each pixel by its standard deviation. Regions in which the algorithm performs poorly are thus suppressed in our analysis. This can be seen in the final panels of Figs. \ref{fig:datareduc_12} and \ref{fig:datareduc} in Appendix \ref{app:datareduc}.

\section{Analysis}
\label{sec:analysis}
\begin{figure}
    \centering
    \includegraphics[width=\linewidth]{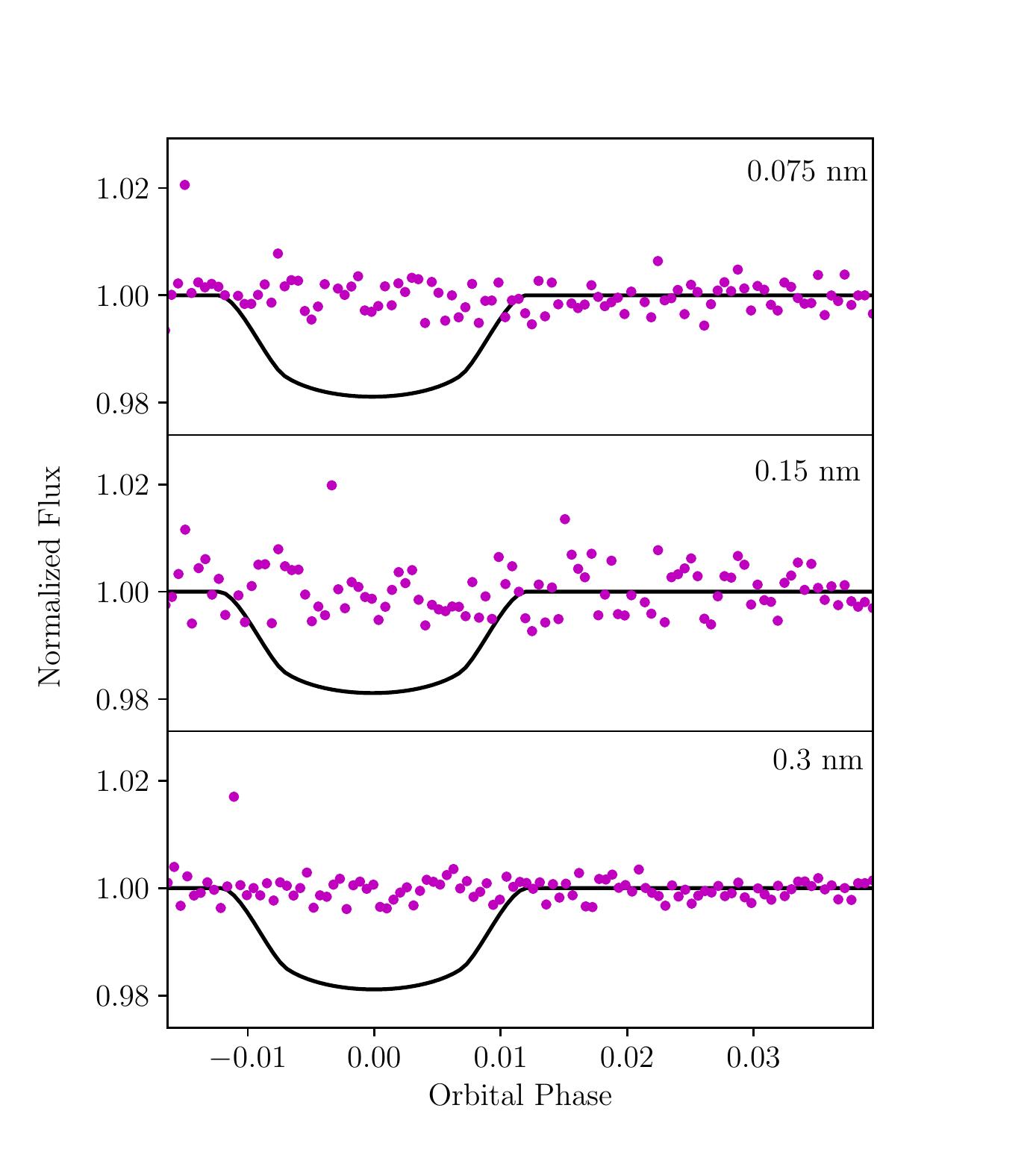}
    \caption{An example of using the \cite{Snellen08} method to analyze the sodium D${}_1$ line in the WASP-69b dataset. The panels represent (from top to bottom) passbands of 0.075, 0.15, and 0.3 nm, and the magenta points correspond to the data while the black line corresponds to a model light curve generated using the methods of \cite{Mandel02}.  We do not detect any signal in any of the three passbands.}
    \label{fig:lightcurve}
\end{figure}

Our initial analysis of the data involved generating light curves as in \cite{Snellen08}. Following their methodology, we integrated the flux within narrow (0.075 nm), medium (0.15 nm), and wide (0.3 nm) passbands around absorption lines of interest to create light curves and look for a signal during the in-transit portion of the data. This was done for both exoplanets in order to search for absorption due to sodium (using the narrow, medium, and wide passband sizes noted above centered on the D${}_1$ and D${}_2$ lines of the sodium doublet, located at 589.592 nm and 588.995 nm respectively) and potassium (again using the narrow, medium, and wide passbands noted above centered on the absorption feature at 769.896 nm). We compared the generated light curves with model light curves made using the methods described in \cite{Mandel02}, as well as the parameters described in Tables \ref{tab:hatp12b} and \ref{tab:wasp69b}. In both bases, we do not detect any signal; an example is shown in Fig. \ref{fig:lightcurve} where we present our analysis of the D${}_1$ absorption line in the WASP-69b data.

However, We note that although our light curves were generated using relatively narrow bins, features may have been obscured or diluted by clouds in the exoplanets' atmospheres. To this end, we chose to subsequently employ the Doppler cross-correlation method that was first successfully used by \cite{Snellen10} to detect carbon monoxide in the atmosphere of HD 209458b, and later used in numerous other works; for example, to detect water in \citealt{Brogi14, Brogi16, Brogi18, Birkby13, Birkby17}, among others. The method involves cross-correlating absorption models with our spectra at a range of Doppler shifts, and then phase-folding the correlation from the in-transit frames and summing over each velocity in order to obtain a map of correlation strength as a function of systemic and Keplerian velocities.

\subsection{Atmospheric Models}
\label{subsec:models}
\subsubsection{Alkali Metals + Water Models}
Model atmospheres for planets WASP-69b and HAT-P-12b were generated using the methods described in \citet{Fortney05,Fortney08}.  Briefly, under the assumption of chemical equilibrium for solar abundances, a model atmosphere was generated, with the atmospheric temperature structure and chemical abundances arrived at iteratively, given models for the incident fluxes of each planet's parent star, and the assumption of radiative-convective equilibrium.  The models represent planet-wide average conditions.  The model's atmospheric opacities, updated since \citet{Fortney08}, include \citet{Barber06} for water and \citet{Allard16} for the alkali metals.

The high-resolution transmission spectra for the planets were generated using the line-by-line code described in \citet{Morley17}.  The transmission spectrum code uses the atmospheric temperature profile, atmospheric abundances, and opacities as used in the generation of the model atmosphere.  Spectra were calculated at a resolution of $R=500,000$.

These models are presented in Fig. \ref{fig:models-j}.

\subsubsection{Water-Only Models}
Transmission spectra of WASP-69b and HAT-P-12b were computed with the line-by-line, plane-parallel radiative transfer code utilised for past work on CRIRES data (e.g. \citealt{deKok14}). The code takes as input a prescribed $T$-$p$ profile from \cite{Parmentier14} and corresponding equilibrium abundances from \cite{Kempton17}, as calculated through the NASA Planet Spectrum Generator (\citealt{Villanueva18}). The radiative transfer calculations account for H$_2$-H$_2$ collision-induced absorption (\citealt{Borysow01, Borysow02}) and molecular opacity from 157,000 water vapour lines extracted from HITEMP 2010 (\citealt{Rothman10}) and accounting for all the main lines contributing to the transmission spectrum, plus additional weak lines significantly contributing to the pseudo-continuum. The choice of HITEMP 2010 has led to several detections of water vapour in high-resolution observations of exoplanet atmospheres and is therefore adopted as a benchmark against which other molecular databases can be compared. The radiative transfer is computed across 50 layers of the planet's atmosphere equally spaced in $\log(p)$, and it accounts for the slanted geometry during transit. Additional input parameters of the models are the stellar and planet radii, and planet surface gravity.

These models are presented in Fig. \ref{fig:models-m}.

\subsection{Doppler Cross-Correlation for Alkali Metal Absorption Features}
\label{subsec:doppler}
In this section, we describe our search for sodium and potassium absorption features in the atmospheres of HAT-P-12b and WASP-69b. The methodology is similar for both elements in both atmospheres.

We correlated each spectrum around a particular absorption line (either potassium or sodium) with models that had been Doppler-shifted to a range of radial velocities between -100 km/s and +100 km/s, with a 1 km/s velocity step between each model. At this point, a particularly strong planetary absorption feature would be present in the correlation map as a diagonal line with a slope corresponding to the changing radial velocity of the planet through its transit. On their own, however, the signals from HAT-P-12b and WASP-69b were not strong enough in a single transit to be detected in the correlation map.

To search for a signal from the exoplanet, we therefore phase-folded the correlation signal from the in-transit spectra of the observation. In-transit spectra were determined through the creation of a model light curve generated with the \texttt{occultquad} package by \cite{Mandel02} and using limb-darkening parameters obtained from \cite{Claret04}. The parameters used in these calculations are summarized in Tables \ref{tab:hatp12b} and \ref{tab:wasp69b}.

\begin{table}[htbp!]
\centering
\caption{Stellar and planetary parameters for HAT-P-12b used in this analysis.}
\begin{tabular}{c  c  c}
Parameter & Value & Reference \\
\toprule
\footnotesize Spectral Type & K4 & \cite{Hartman09} \\
$R_*$ ($R_\odot$) &  0.701 & \cite{Hartman09} \\
$K_p$\tablenotemark{1} (m/s) & 35.8 & \cite{Hartman09} \\
$T_{14}$\tablenotemark{2} (days) & 0.0974 & \cite{Hartman09} \\
$m_p$ ($M_{\text{Jup.}}$) & 0.201 & \cite{Mancini18} \\
$R_p$ ($R_{\text{Jup.}}$) & 0.919 & \cite{Mancini18} \\
$V_\mathrm{sys}$ (km/s) & -40.4589 & \cite{Mancini18} \\
\footnotesize Orbital Period (days) & 3.21305992 & \cite{Mancini18} \\
\footnotesize Mid-Transit (BJD) &  \footnotesize 2455328.49068 & \cite{Mancini18} \\
\footnotesize Semi-Major Axis (AU) & 0.0384 & \cite{Hartman09} \\
\footnotesize Inclination (deg.) & 89.10 & \cite{Mancini18} \\
R${}_p$/R${}_*$ & 0.13898 & \cite{Mancini18} \\
a/R${}_*$ & 11.77 & \cite{Hartman09} \\
$\mu_1$ & 0.6508 & \cite{Claret04} \\
$\mu_2$ & 0.1184 & \cite{Claret04} \\
\hline
\label{tab:hatp12b}
\end{tabular}
\vspace*{-6mm}
\flushleft
\tablenotetext{1}{$K_p$: radial velocity semi-amplitude}
\tablenotetext{2}{$T_{14}$: total transit duration, time between first and last contact}
\end{table}

\begin{table}[htbp!]
\centering
\caption{Stellar and planetary parameters for WASP-69b used in this analysis.}
\begin{tabular}{c  c  c}
Parameter & Value & Reference \\
\toprule
\footnotesize Spectral Type & K5 & \cite{Anderson14}  \\
$R_*$ ($R_\odot$) &  0.813 & \cite{Anderson14} \\
$K_p$\tablenotemark{1} (m/s) & 38.1   & \cite{Anderson14}  \\
$T_{14}$\tablenotemark{2} (days) & 0.0929  & \cite{Anderson14}  \\
$m_p$ ($M_{\text{Jup.}}$) & 0.260  & \cite{Anderson14}  \\
$R_p$ ($R_{\text{Jup.}}$) & 1.057 & \cite{Anderson14}  \\
$V_\mathrm{sys}$ (km/s) & -9.62826 & \cite{Anderson14} \\
\footnotesize Orbital Period (days) & 3.8681382 & \cite{Anderson14} \\
\footnotesize Mid-Transit (BJD) & \footnotesize 2455748.83344 & \cite{Anderson14} \\
\footnotesize Semi-Major Axis (AU) & 0.04525 & \cite{Anderson14} \\
\footnotesize Inclination (deg.) & 86.71 & \cite{Anderson14} \\
R${}_p$/R${}_*$ & 0.134 & \cite{Anderson14} \\
a/R${}_*$ & 11.97 & \cite{Anderson14} \\
$\mu_1$ & 0.5503 & \cite{Claret04} \\
$\mu_2$ & 0.2012 & \cite{Claret04} \\
\hline
\label{tab:wasp69b}
\end{tabular}
\vspace*{-6mm}
\flushleft
\tablenotetext{1}{$K_p$: radial velocity semi-amplitude}
\tablenotetext{2}{$T_{14}$: total transit duration, time between first and last contact}
\end{table}

Next we interpolated the correlation signal to the rest frame of the planet and summed over time in order to increase the strength of the signal. Since the planetary radial velocity semi-amplitude is not known a priori, we interpolated to values ranging from 1 km/s to 200 km/s, with a 0.5 km/s step between each value. The end result is a correlation map over a range of systemic velocities ($V_\mathrm{sys}$) and planetary radial velocity semi-amplitudes ($K_p$). If a feature is present in the planet's atmosphere, we expect to see it as a strong correlation signal at the planetary radial velocity semi-amplitude $K_p$ and a systemic velocity of $V_\mathrm{sys} = 0$ km/s. 

The results of this process are presented in section \ref{sec:analysis} and Fig. \ref{fig:modelinjections}. \newline

\subsection{Doppler Cross-Correlation for Water Absorption Features}
\label{subsec:dopplerwater}
Whereas alkali metals produce only a few absorption lines, there are thousands of absorption lines due to water spanning multiple orders in the data. The methodology used to search for water is similar to that described above in Section \ref{subsec:doppler}; however, rather than correlating over a small window including the relevant absorption lines, we correlate each order of the data with the corresponding wavelength range in the model, and sum the resulting correlation maps. We avoid the orders containing sodium and potassium (orders 3 and 11 for HAT-P-12b and 7 and 16 for WASP-69b), as these are known to be strong absorption features and would dominate the correlation.

\begin{figure*}
\includegraphics[width=\textwidth]{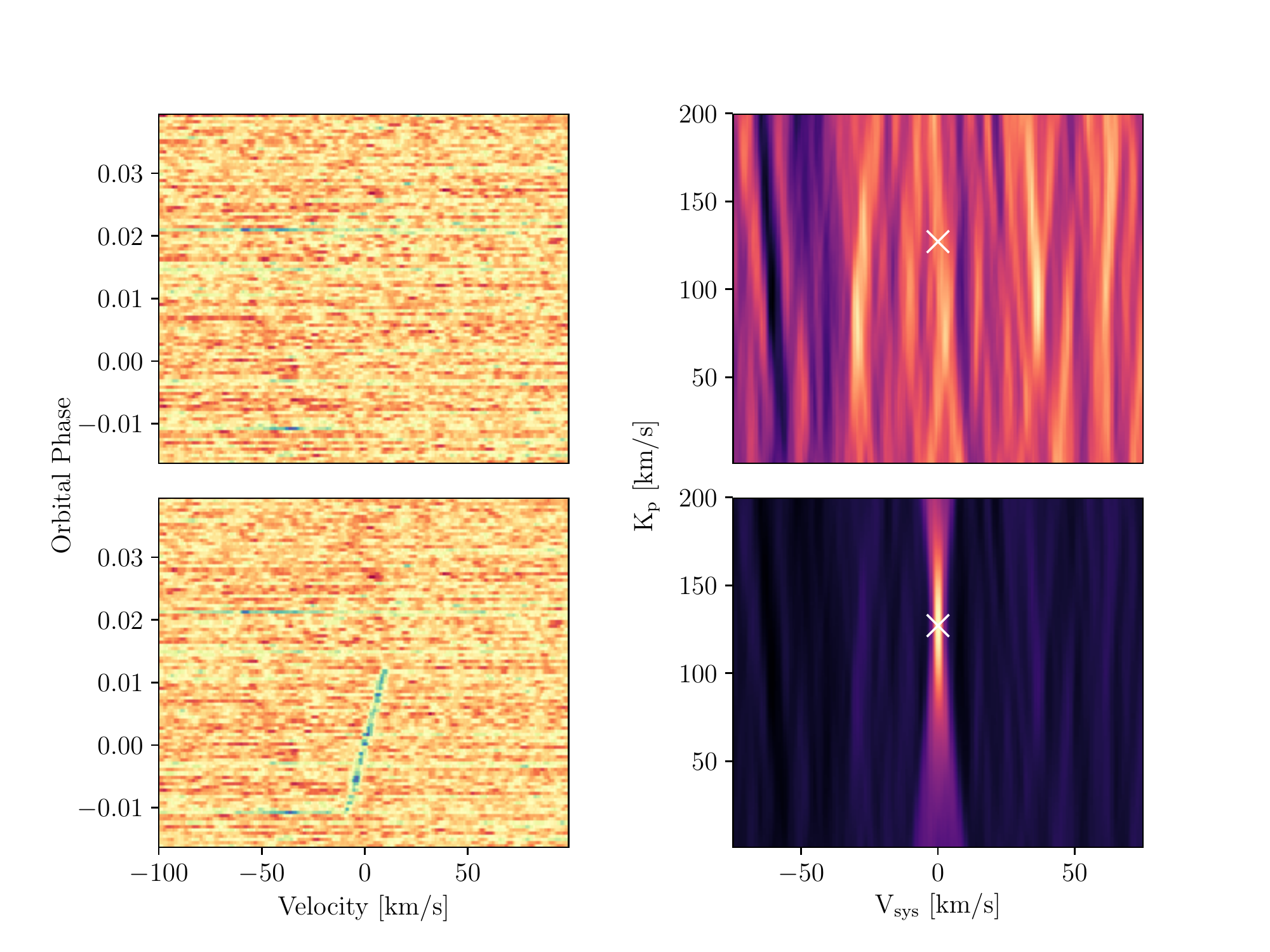}
\caption{Example correlation maps (left panels) and phase-folded correlations (right panels) obtained from cross-correlating the thousands of water lines across each spectrum of our WASP-69b observations with models at Doppler shifts of -100 km/s to +100 km/s. The colour represents the correlation strength. The top panels show the analysis process for our data, and the bottom panels show the same process repeated on the data with a water model injected (see Fig. \ref{fig:models-m}). The signal from water is visible in the bottom-left panel as a diagonal line across the in-transit portion of the map, corresponding to roughly -0.01 to + 0.01 orbital phase. Likewise, the signal is visible at the Keplerian velocity of the exoplanet in the bottom-right panel, at the expected V${}_{\text{sys}}$ and K${}_{p}$ of the planet: 0 km/s and $\sim$ 127 km/s respectively. The cross indicates this position in both panels. The fact that no similar signal is visible in the original analysis of our data indicates that water is not present in the atmosphere of WASP-69b at the strength of the injected model.}
\label{fig:wasp69xmatteo}
\end{figure*}

A detailed example of the process is presented in Fig. \ref{fig:wasp69xmatteo}, and the results of our analysis are presented in section \ref{sec:analysis} and Fig. \ref{fig:water}.

\subsection{Model Injection/Recovery Tests}
To determine the significance of our results and detection limits of our observations, we performed model injection/recovery tests for several absorption features at varying strengths for each planet. This was done by multiplying the in-transit frames of the raw data (i.e. the data processed through the reduction piplines available at the telescopes but before any further reduction) by an atmospheric model in the planet's reference frame. We used the atmospheric models described in Section \ref{subsec:models}, but varied the model strength throughout the injection/recovery tests. In this way, we are able to determine the minimum strength at which the model would be recovered by our analysis, and can use this to constrain the atmospheric properties of the planets.

After injecting the models, we followed the data reduction processes described in Section \ref{sec:reduc} and carried out the Doppler cross-correlation technique described in Section \ref{subsec:doppler}. 

\subsection{Detection Significance}
If a candidate signal does not surpass $3\sigma$, we do consider it to be significant. If a candidate signal \textit{does} surpass $3\sigma$, we consider the candidate as warranting further investigation. As will be discussed further in section \ref{subsec:hatp12b}, we caution that noise in the data can lead to features approaching the $3\sigma$ level, and thus any potential signals must be treated carefully.

In order to determine the significance of our results, we followed the methods described in \cite{Esteves17}. This involved creating $1\sigma$ and $3\sigma$ confidence levels for each feature in each planet by randomly selecting 24 frames in the case of HAT-P-12b and 58 frames in the case of WASP-69b (corresponding to the total number of in-transit frames for each dataset), assigning each an in-transit phase, and carrying out the correlation and phase-folding as described in Section \ref{subsec:doppler}. The process was repeated 10,000 times in order to determine $1\sigma$ and $3\sigma$ confidence levels. 

\section{Results and Discussion} \label{sec:discussion}
In this section, we present the results of applying the Doppler cross-correlation method as described in Sections \ref{subsec:doppler} and \ref{subsec:dopplerwater} to our observations. In particular, we are looking for features caused by atmospheric absorption due to sodium, potassium, and water.
\subsection{HAT-P-12b}\label{subsec:hatp12b}
Our results are summarized in Fig. \ref{fig:modelinjections}. As can be seen in the top row of Fig. \ref{fig:modelinjections}, there is a peak at $3.2\sigma$ in the data at a systemic velocity of 0 km/s and a planetary orbital velocity of $\sim$ 130 km/s, consistent with a signal due to atmospheric sodium absorption. This feature has not been observed in any previous analyses of HAT-P-12b (see, e.g., \citealt{Line13} and \citealt{Sing16}).

In the case of potassium, we observe a peak in the phase-folded correlations at V${}_{\text{sys}} = 0$ outside of the $1\sigma$ contour, but note that it does not exceed the $3\sigma$ confidence level. We also note, however, that there are many additional peaks at $> 1\sigma$ that are likely due to noise in the data. A similar phenomenon was observed by \cite{Esteves17} in their analysis of 55 Cnc e's atmosphere, and they found that repeating the Doppler cross-correlation with a pure white noise spectrum resulted in similar features in the phase-folded correlations.

Based on the model injections, we can rule out potassium in the atmosphere of HAT-P-12b down to an amplitude of 2\% relative to the normalized flux. Any models injected lower than this are not detected beyond the $3\sigma$ level in our analysis.

In the case of water, our results are presented in Fig. \ref{fig:water}. As in Fig. \ref{fig:modelinjections}, 1- and $3\sigma$ confidence levels are shown, as well as phase-folded correlations for the data alone and the data with an atmospheric model injected. We make use of two separate models (see Section \ref{subsec:models}): one that includes a full atmospheric treatment of each planet, and one that only includes the signal due to water in the exoplanetary atmosphere. The results for HAT-P-12b are in the top row of Fig. \ref{fig:water}. An injection of the first model yields a detection outside the $3\sigma$ level, indicating that water is not present in the atmosphere of HAT-P-12b at this model strength.

Previous analyses of HAT-P-12b's atmosphere have revealed a cloudy spectrum with very little detectable atmospheric absorption (\citealt{Line13, Sing16, Alexoudi18}), which is consistent with our observations. \cite{Sing16} and \cite{Alexoudi18} \textit{did} detect evidence of weak potassium absorption in the planet's atmosphere, and although we did not confidently observe the same, we note that this could be due to the noise in our data or the fact that only one planetary transit was used in our analysis, resulting in a lower signal-to-noise ratio than would be required to confidently detect atmospheric potassium.

\subsection{WASP-69b}
\begin{figure*}[htbp!]
\centering
\includegraphics[width=\textwidth]{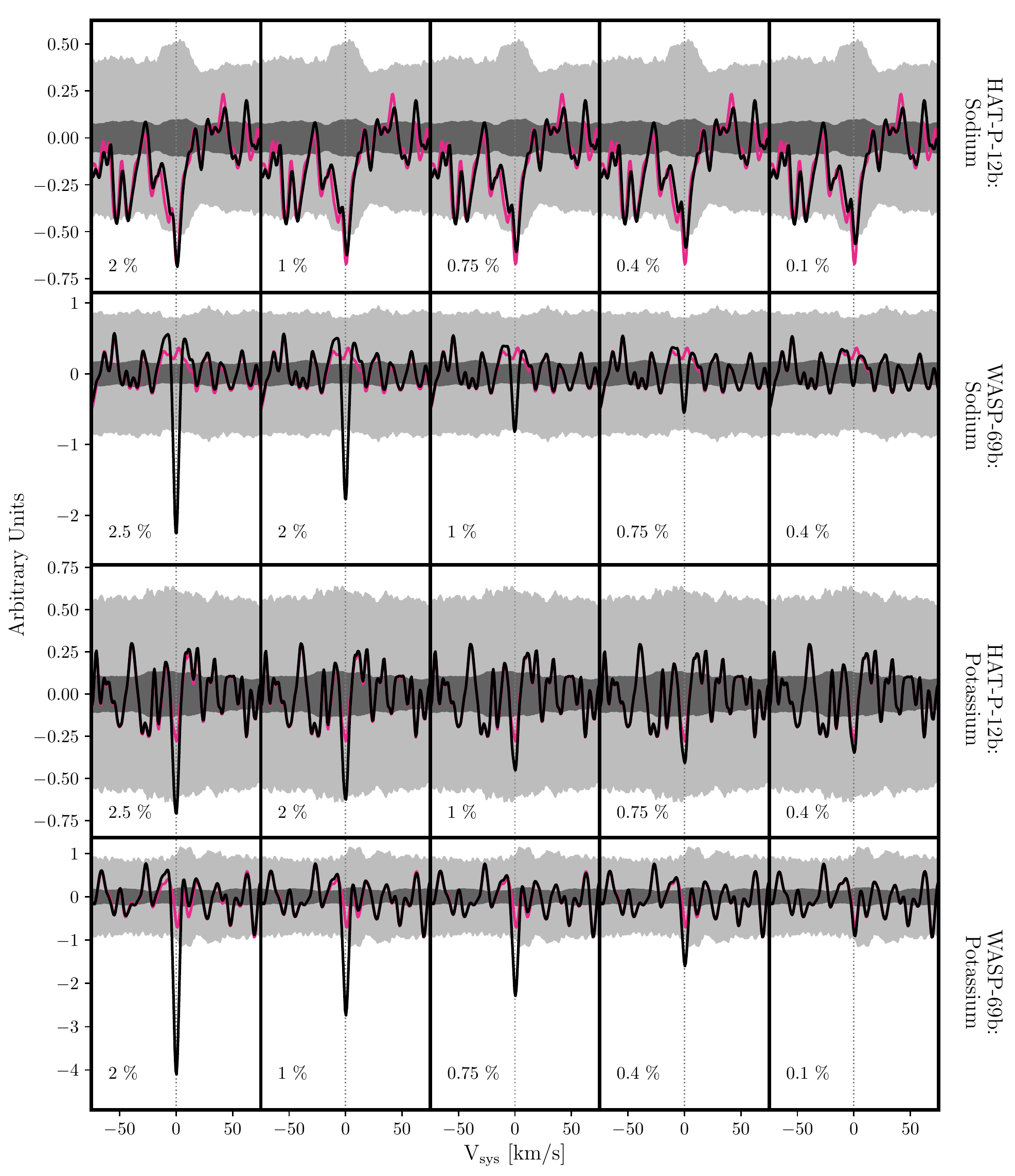}
\caption{The results of injecting models of various strengths into our data, and repeating the Doppler cross-correlation process. In all panels, the magenta line represents the original data and the black line represents the data with a model injected, where the strength of the injected model is indicated in the bottom-left corner of the panel. We note that this percentage refers to a percentage relative to the normalized flux, i.e. a percentage of the stellar surface area. The dark grey contour represents a $1\sigma$ confidence level, while the lighter grey contour represents a $3\sigma$ confidence level. The relevant planet and element are indicated on the right-hand side of each row. For HAT-P-12b we phase-fold the data to the planet's orbital radial velocity K${}_p = 130$ km/s, and for WASP-69b we phase-fold the data to the planet's orbital radial velocity K${}_p = 127$ km/s. We also take into account the orbital systemic velocity of each system; see Tables \ref{tab:hatp12b} and \ref{tab:wasp69b}.}
\label{fig:modelinjections}
\end{figure*}

As in Section \ref{subsec:hatp12b}, the results of cross-correlating our original and injected data with model atmospheric spectra for various absorption features can be seen in Fig. \ref{fig:modelinjections}. In the case of both sodium and potassium, there are no significant (i.e. $> 3\sigma$) peaks in the WASP-69b data at a systemic velocity of 0 km/s. For the sodium correlations (row 2 in Fig. \ref{fig:modelinjections}), we do not observe any features at V${}_{\text{sys}} = 0$, and the majority of features present at other systemic velocities are contained within the $1\sigma$ contour. The data with a model injected yields a $3\sigma$ detection down to 1 \% of the flux with respect to the normalized flux, meaning that these model strengths can confidently be ruled out from the planet's atmosphere.

For the potassium correlations (row 4 in Fig. \ref{fig:modelinjections}), we note that there \textit{is} a peak in the data at V${}_{\text{sys}} = 0$ well outside the $1\sigma$ level; however, it does not surpass the $3\sigma$ level. Furthermore, injected models at strengths down to $\sim$ 1\% relative to the normalized flux result in $> 3\sigma$ peaks in the phase-folded correlations. If potassium is present in the planet's atmosphere, it is therefore at a relatively low amplitude and would require more observations or a greater signal-to-noise ratio to observe.

We also note the presence of a feature slightly exceeding 3$\sigma$ at V${}_{\text{sys}} \sim + 70$ km/s in the potassium correlations. We believe that this feature, similar to the many features at $> 1\sigma$ discussed in section \ref{subsec:hatp12b}, is caused by remaining noise/systematics in the data. This reinforces the fact that potential candidate signals (i.e. those surpassing $3\sigma$, such as that detected at V${}_{\text{sys}}=0$ km/s for sodium absorption in HAT-P-12b as discussed in section \ref{subsec:hatp12b}) must be treated with caution.

\begin{figure*}
    \centering
    \includegraphics[width=\textwidth]{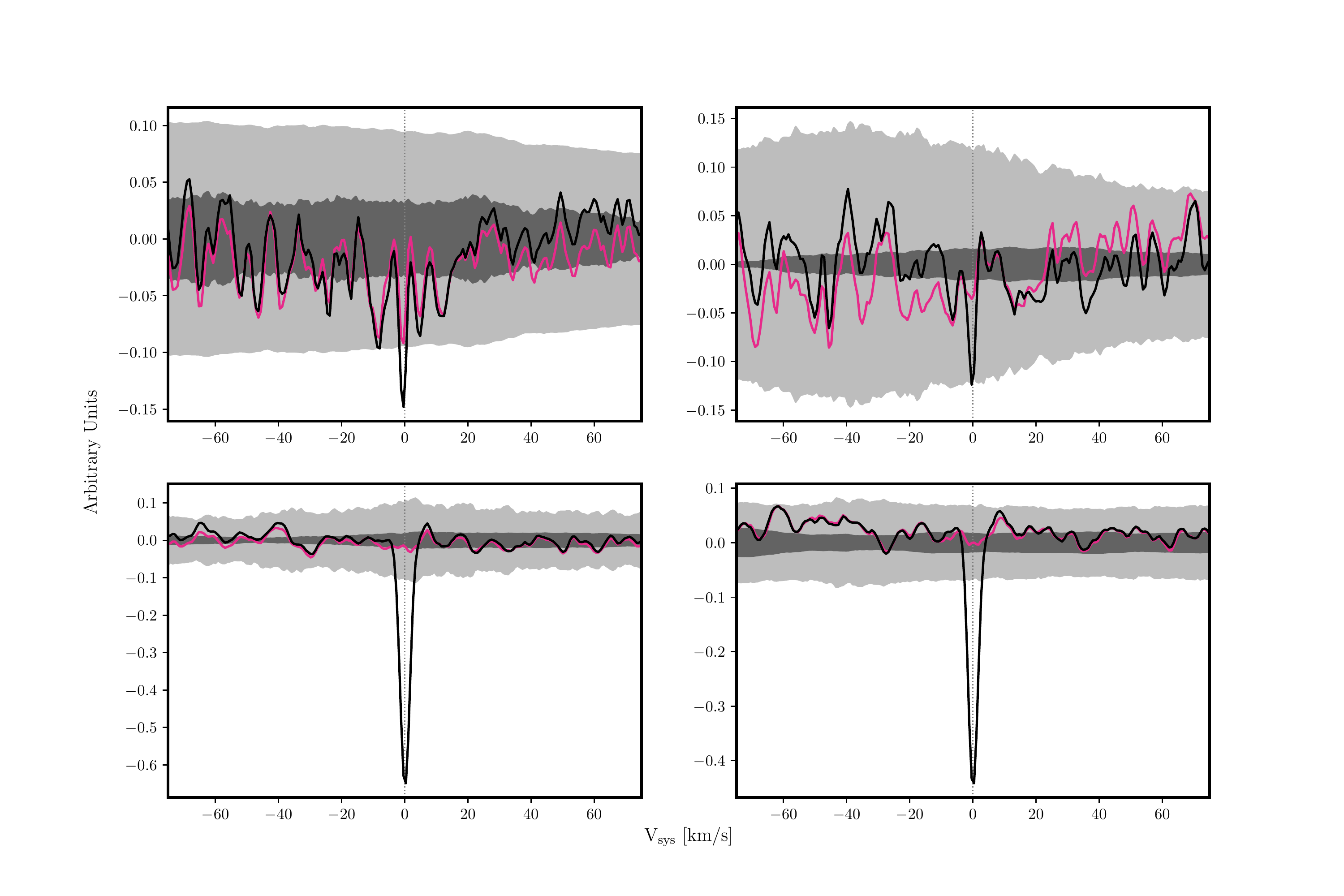}
    \caption{The results of injecting atmospheric absorption models into our data and repeating the Doppler cross-correlation process. In all panels, the magenta line represents the original data and the black line represents the data with a model injected. This process is repeated for each order of the data (aside from those containing strong absorption features due to sodium and potassium). The top row shows our HAT-P-12b observations and the bottom our WASP-69b observations; and the left column shows injections with the models containing alkali metals and water, while the right shows injections with the models containing only water.}
    \label{fig:water}
    \includegraphics[width=\textwidth]{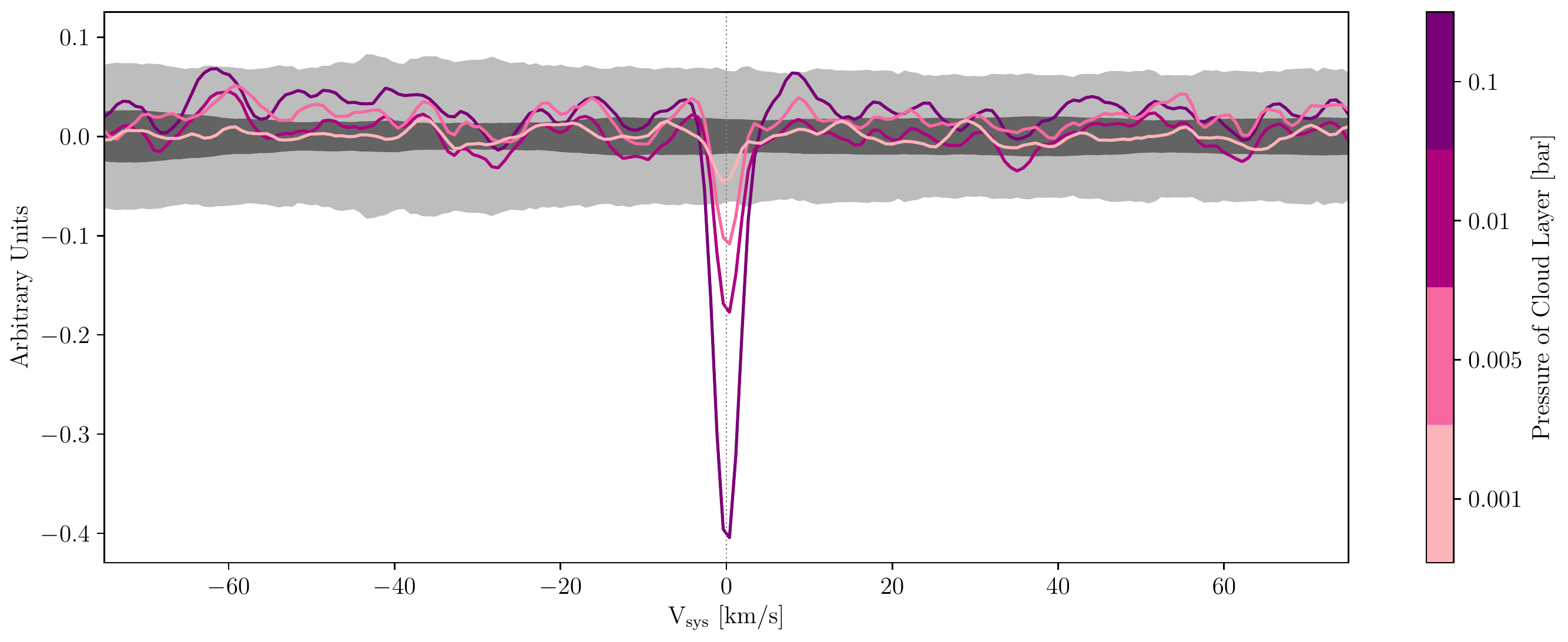}
    \caption{The results of injecting model spectra with cloud layers at various pressures into our WASP-69 data. The pressures corresponding with each layer are indicated in the colourbar. This analysis was performed using the water-only models. We see that at a pressure of 1 mbar, the injected signal is no longer detectable in our data.}
    \label{fig:clouds}
\end{figure*}

Our analysis of water absorption features is presented in Fig. \ref{fig:water}. In both cases, the analysis of the data with an injected model results in a strong signal well outside the $3\sigma$ range, whereas the signal from the data itself does not surpass this level. Thus we conclude that water is not present in the atmosphere of WASP-69b at the strength of either model. Furthermore, the strong signal that is detected when water models are injected into the WASP-69b data allows us to investigate the presence of cloud decks in the exoplanet's atmosphere. Following the analysis presented in \cite{Pino18}, we inject models with a simulated cloud layer into the data,  allowing us to place a limit on the altitude at which clouds might exist in the atmosphere of WASP-69b. We choose various cloud pressures and corresponding transit depths in our models, and repeat the cross-correlation after cutting off the model below this depth. In Fig. \ref{fig:clouds}, we show injections of models with simulated cloud layers at pressures of 0.1 bar, 0.01 bar, 5 mbar, and 1 mbar. Down to a pressure of 5 mbar the signal is still detected outside the $3\sigma$ confidence level; however, a cloud layer at a pressure of 1 mbar does not surpass this level and therefore cannot be ruled out in the atmosphere of WASP-69b.

The atmosphere of WASP-69b has previously been studied by \cite{Casasayas17} with the medium-resolution HARPS-North spectrograph. In contrast with our analysis, their observations led to a $5\sigma$ detection of atmospheric sodium absorption in just the D${}_2$ line of the sodium doublet, at 588.995 nm (\citealt{Casasayas17}). In order to better compare our analysis with theirs we repeated the Doppler cross-correlation technique as well as the model injection/recovery tests with each line of the sodium doublet separately; however, this did not significantly change our results. In particular, we did not observe a strong correlation signal when carrying out our analysis in just the D${}_2$ line of the sodium doublet.

\subsection{The Rossiter-McLaughlin Effect and Centre-to-Limb Variations}
\label{subsec:rm-clv}
Here we consider the possible consequences of the Rossiter-McLaughlin (RM) effect and centre-to-limb variations (CLV) on our analysis.

The consequences of the RM effect/CLV have been addressed in previous work in the field. In particular, centre-to-limb variations of stellar lines may affect transmission spectra (\citealt{Yan17}), while the RM effect can introduce misalignments or spurious detections/non-detections into final results (\citealt{Louden15, Barnes16}). 
However, we do not think that the RM effect and CLV will have had a noticeable impact on our results. In particular, we note that HAT-P-12 is an extremely slow rotator, with $v\sin i_* = 0.5 \pm 0.5$ km/s, as determined in \cite{Mancini18} using HARPS-N measurements. Furthermore, measurements of the sky-projected orbital obliquity angle $\lambda$ are unconstrained, but may be consistent with spin-orbital alignment ($\lambda = -54^{+41}_{-13} { }^{\circ}$).

WASP-69 is a slightly faster rotator, with $v\sin i_* = 2.2 \pm 0.4$ km/s (\citealt{Anderson14}). In their analysis using the HARPS-N spectrograph, \cite{Casasayas17} model the RM effect of WASP-69b. They derive a value of $\lambda = 0.4^{+2.0}_{-1.9} { }^{\circ}$ for the sky-projected orbital obliquity angle. It is found that the RM effect does not significantly impact their results.

\section{Conclusion}
\label{sec:conclusion}
We have presented our analysis of high-resolution optical transmission spectroscopy of two sub-Saturn mass transiting exoplanets, HAT-P-12b and WASP-69b. While the majority of high-resolution, ground-based efforts to study exoplanetary atmospheres have targeted massive planets, HAT-P-12b and WASP-69b represent an unexplored area of parameter space, with lower masses and cooler temperatures than any hot Jupiters which have previously had their atmospheres characterized via ground-based transmission spectroscopy.

In this paper, we presented the results of the Doppler cross-correlation technique, which takes advantage of the high resolution of our observations and the large radial velocities of our exoplanets in order to correlate sophisticated model spectra with our data in the search for atmospheric absorption features. In particular, we targeted sodium and potassium, two alkaline absorption features which should extend high into the exoplanets' atmospheres. We also used similar techniques to search for water, which is made up of thousands of absorption lines in the wavelength range of our observations.

Our analysis revealed a $3.2\sigma$ detection of atmospheric sodium in HAT-P-12b, which has not been detected in any previous analyses of this exoplanet's atmosphere. 

\acknowledgments

Author EKD is supported in part by an Ontario Graduate Scholarship and an NSERC CGS-M. We thank Abhinav Jindal, Ryan Cloutier, and Lisa Esteves for helpful discussions. The authors acknowledge the role that Roxana Lupu, Richard Freedman, Didier Saumon, and Caroline Morley played in curating the opacity database used in some of these model spectrum calculations.

\clearpage
\appendix
\section{Data Reduction}
\label{app:datareduc}

\begin{figure}[htbp!]
\centering
\includegraphics[width=\textwidth]{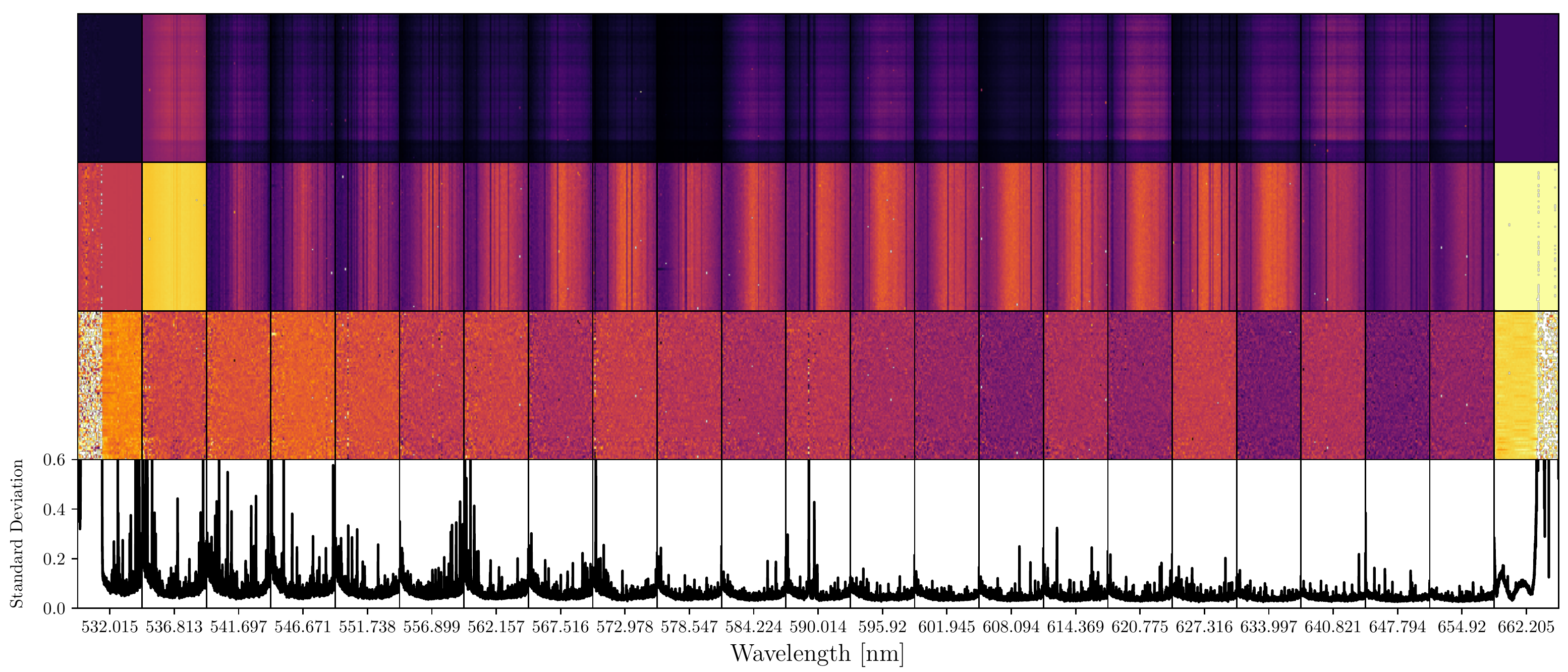}
\caption{The full data reduction process as applied to the HAT-P-12b observations from the blue CCD. The top panel shows the raw data as they are received after the initial reduction pipeline at the telescope (see Section \ref{sec:reduc}). The second panel shows the results of the blaze correction and cosmic-ray filtering (see Section \ref{sec:reduc}). The third panel shows the data after five iterations of {\sc Sysrem} have been applied, and the bottom panel shows the standard deviation along each wavelength band of the third panel.}
\label{fig:datareduc_12}
\end{figure}

\begin{figure}[htbp!]
\centering
\includegraphics[width=\textwidth]{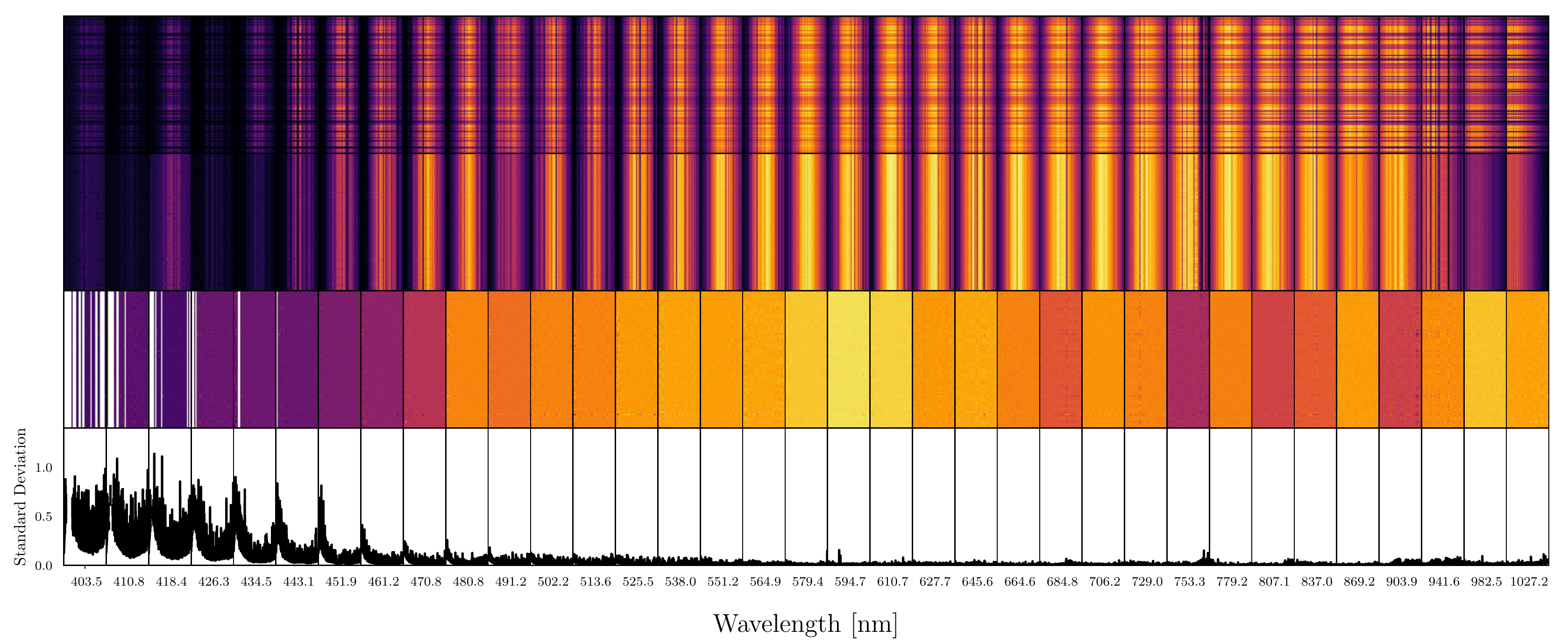}
\caption{The full data reduction process as applied to the WASP-69b observations. Explanations for each panel of the figure are provided in the caption of Fig. \ref{fig:datareduc_12}.}
\label{fig:datareduc}
\end{figure}

\vfill
\newpage

\section{Models}
\label{app:models}

\begin{figure}[htbp!]
\centering
\includegraphics[width=0.95\textwidth]{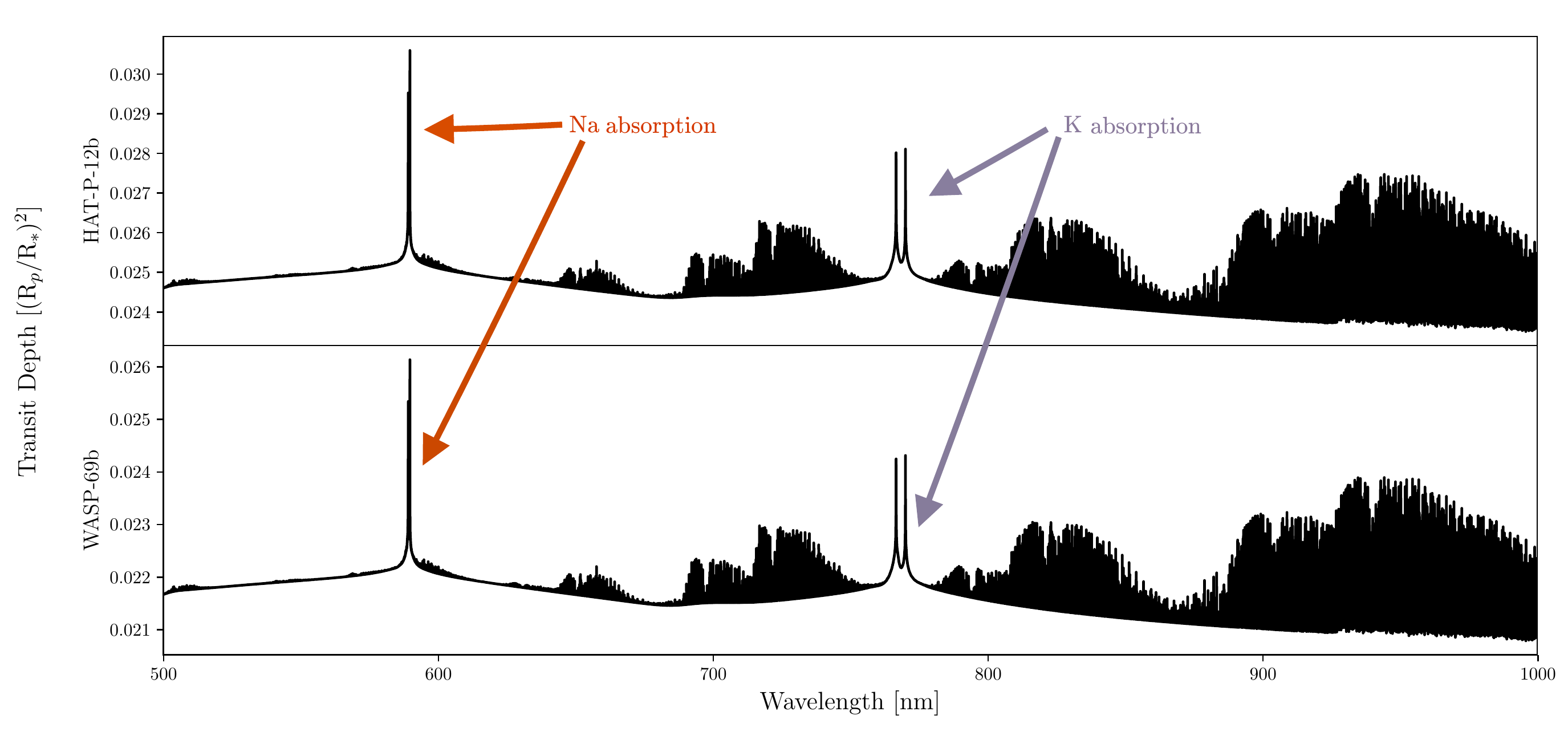}
\caption{The model spectra used in our analysis. The top panel shows the spectrum calculated for HAT-P-12b, and the bottom panel shows the spectrum calculated for WASP-69b. Water bands can be identified from Fig. 11, but the locations of other important species are marked.}
\label{fig:models-j}
\end{figure}

\begin{figure}[htbp!]
\centering
\includegraphics[width=0.95\textwidth]{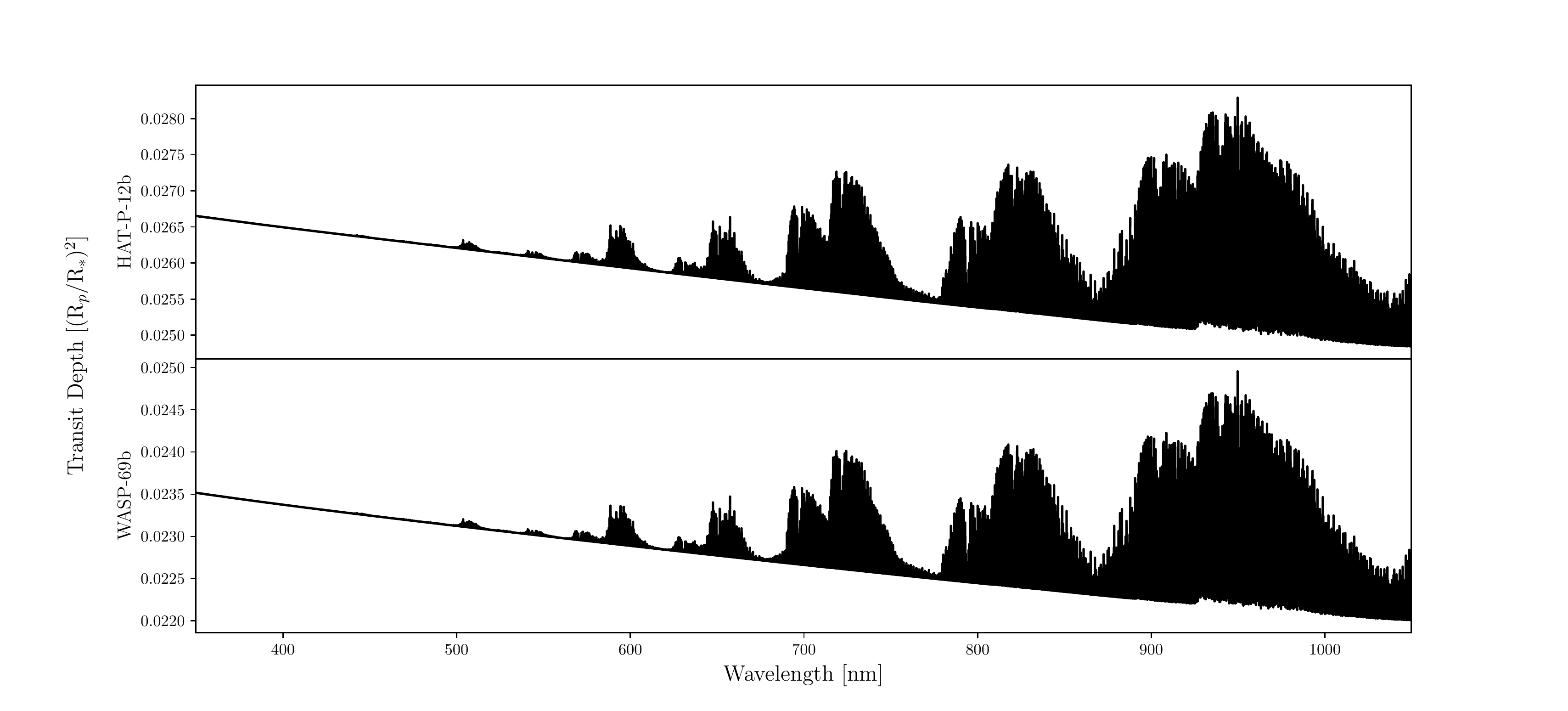}
\caption{The model spectra used in our analysis. The top panel shows the spectrum calculated for HAT-P-12b, and the bottom panel shows the spectrum calculated for WASP-69b. These models only include lines from water.}
\label{fig:models-m}
\end{figure}

\end{document}